\documentclass[12pt]{JHEP}
\usepackage{epsfig}

\def\qrd{\buildrel{\scriptscriptstyle{(-)}}\over{{q}}}

\title{Gluon Contributions to Parity-Violating Asymmetries in Polarized
Proton-Proton Scattering }
\author{John Ellis, Stefano Moretti and Douglas A. Ross\thanks{On leave
of absence from:
 Department of Physics and Astronomy, University of Southampton,
 Southampton SO17 1BJ, U.K.}
 \\ Division Th\'{e}orique, CERN, 1211
 Gen\`{e}ve 23, Switzerland}

\abstract{
We report on a calculation of one-loop weak corrections to polarized
quark-gluon scattering and the corresponding crossed channels. Such
contributions are suppressed formally by one power of $\alpha_s$ relative
to $W-$ or $Z-$mediated quark-quark scattering, but would enable the spin
asymmetry of the gluon distribution to contribute to parity-violating
asymmetries that will soon be investigated in polarized proton-proton
scattering experiments at RHIC.  In certain kinematic regions, gluon
contributions to parity-violating asymmetries can be as large as
10\% of the tree-level $W-$ and $Z-$exchanges in quark-quark scattering,
but usually only where the parity-violating asymmetries are already small. 
}

\preprint{hep-ph/0102340\\
CERN-TH/2001-053\\
February 2001}

\def\beq{\begin{equation}}
\def\eeq{\end{equation}}
\def\bea{\begin{eqnarray}}
\def\eea{\end{eqnarray}}

\def\fns{\left[  \ln\left(1-\frac{s}{m^2}\right) 
  \ln\left(\frac{-t}{m^2}\right) \, + \, 
  {\rm Li}_2\left(\frac{s}{m^2}\right) \right] }

\def\fnu{\left[  \ln\left(1-\frac{u}{m^2}\right) 
  \ln\left(\frac{-t}{m^2}\right) \, + \,
   {\rm Li}_2\left(\frac{u}{m^2}\right) \right]} 

\def\ct0{\left[ \frac{\pi^2}{6} \, - \, {\rm Li}_2\left(1+\frac{t}{m^2} \right)
 \right]}

\begin{document}

\section{Introduction}

The forthcoming generation of hadron-hadron colliders - the Tevatron with
Run 2, RHIC with polarized proton beams (RHIC-Spin) and the LHC - will
produce unprecedented data on high-$p_T$ jets that merit the most accurate
theoretical calculations.  Next-to-leading order (NLO) jet calculations in
QCD have been available for some time \cite{ellsex}, whereas the major
task of providing the complete NNLO QCD result is about to be
completed~\cite{3jet,glover}. Formally, the latter introduce corrections
relative to the leading-order predictions that are of order $(\alpha_s /
\pi)^2 = {\cal O} (10^{-3})$, with the likelihood of parametric
enhancements (or suppressions).  When computing to this accuracy, one may
then wonder whether NLO electroweak (EW) corrections could also be
important. In fact, these are formally of order $(\alpha_W/ \pi) = {\cal
O} (10^{-2})$, again with the likelihood of parametric enhancement or
suppression. 

We report in this paper on a pilot one-loop calculation of purely weak
corrections to high-$p_T$ jet production.  We concentrate on quark-gluon
scattering and processes related to this by crossing, namely $\qrd +~g \to
\qrd +~g$, $q~ +~ \bar{q} \to g~ +~ g$ and $g~ +~ g \to q~ +
\bar{q}~$\footnote{Hereafter, the notation $q$ is intended to represent any
possible quark flavour.}.  In order to focus on a possible distinctive
experimental signal, we discuss parity-violating asymmetries in these
processes that are in principle measurable at RHIC-Spin~\cite{Bunce}.
These weak corrections provide, in principle, sensitivity to the
helicity-dependent component of the gluon distribution function inside a
polarized proton, $\Delta G$, that is instead absent in pure QCD
\cite{PDFs}.

Contributions to such parity-violating asymmetries already arise at tree
level from $q + q \to q + q$ and $q +\bar{q} \to q + \bar{q}$ scattering
processes, due to purely weak amplitudes as well as to the interference of
the latter with the QCD ones. These processes are sensitive to the
helicity-dependent components of the quark distributions, $\Delta Q_q$,
and their effects on parity-violating asymmetries are well known and
rather small, at the level of few percent at best~\cite{bgs}.  As we show
below, the new contributions that we calculate here are even smaller, as
indeed expected on the ground of a na{\"\i}ve estimate based on counting
powers of coupling constants, without any benefit from an enhancement due
to the spin asymmetry of the gluon distribution function.  For plausible
values of the latter, there do exist regions of the high-$p_T$ jet phase
space where the $\Delta G$-dependent contributions to the parity-violating
asymmetries are non-negligible in comparison with the $\Delta
Q_q$-dependent ones, but the experimental sensitivity seems unlikely to be
able to disentangle the two, primarily because of the small absolute
values of the observables. 

Looking on the bright side, our results help stabilise the Standard Model
(SM)  predictions for parity-violating asymmetries at RHIC-Spin, thereby
providing a solid baseline in the search for new physics.  These
parity-violating asymmetries have been proposed as signatures of new
physics, such as a $W'$, $Z'$ or contact interactions that violate parity
\cite{NewPhysics}. In other terms, the claimed RHIC-Spin sensitivities to
these signatures are validated by our results. 

A caveat to our analysis is that we have not computed the one-loop weak
corrections to $q~+ \qrd \to q~+ \qrd $. In fact, for the QCD-mediated
scattering, they are formally of the same order as those for $\qrd +~g \to
\qrd +~g$, $q~+~\bar{q} \to g~+~g$ and $g~+~g \to q~+~\bar{q}$. Since such
processes give rise to parity-violating asymmetries already at tree level
\cite{bgs}, such a computation would amount to a QCD correction to their
effects and is therefore unlikely to have a significant qualitative or
quantitative effect on those asymmetries. 

\section{The one-loop weak corrections to gluon scattering matrix
elements}

\FIGURE{
\centerline{\epsfig{file=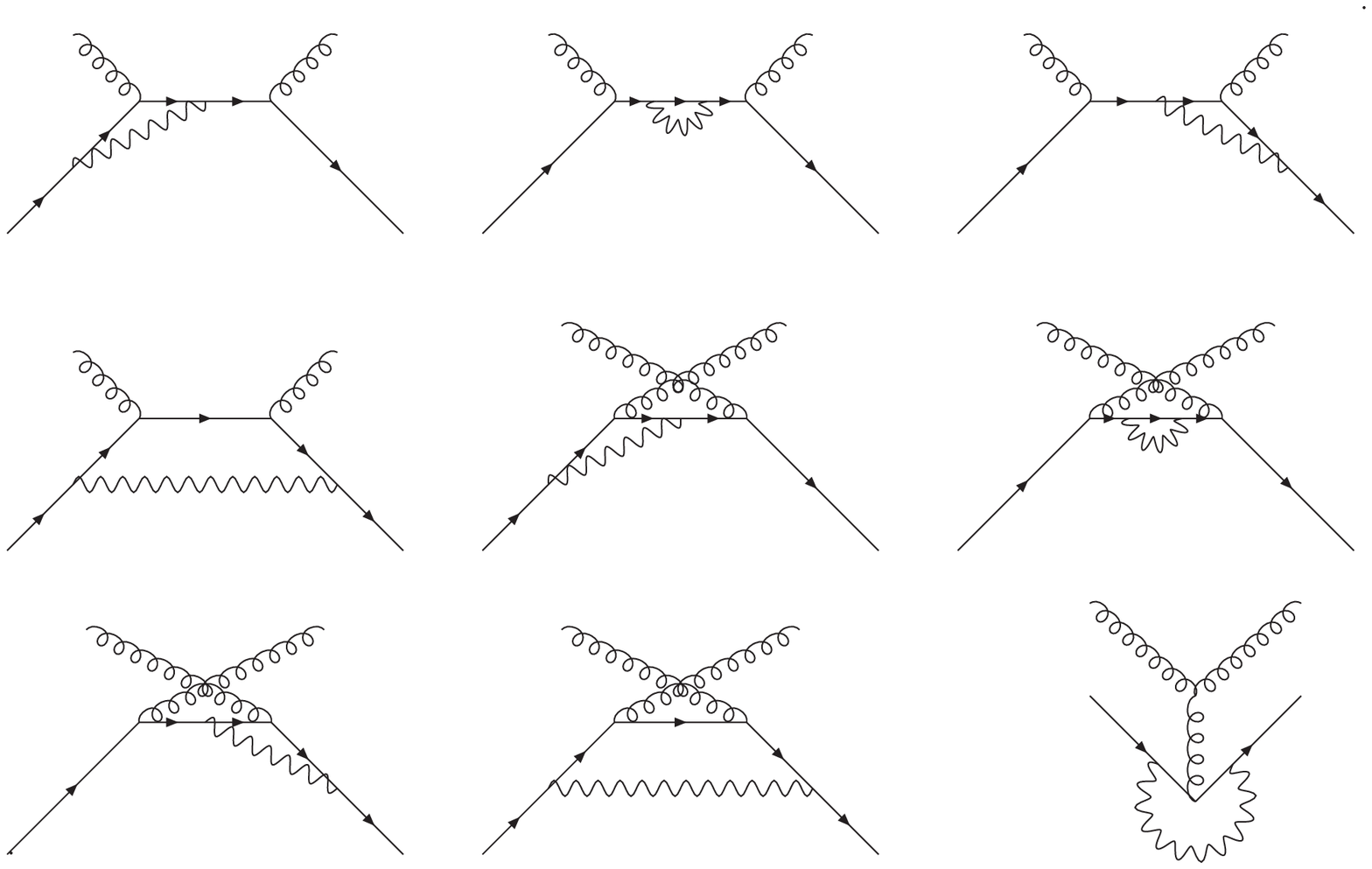, width=14cm}}
\caption{Diagrams contributing to the 
one-loop weak corrections to quark-gluon scattering.
The helical lines represent gluons and the wiggly lines refer to
either a $W-$ or $Z-$boson.}
\label{fig1}
}

One-loop graphs yielding weak corrections to the process $$ \, q + g \,
\to \, q + g \, $$ are shown in Fig.~\ref{fig1}.  In addition to those
shown, there is a correction due to the on-shell fermion wavefunction
renormalization for each tree-level graph, which we have not shown. We
have calculated the graphs in Fig.~\ref{fig1} with the help of FeynCalc
\cite{feyncalc} and FORM \cite{form}. 

The tree-level amplitudes ${\cal A}_{\lambda\lambda',\sigma}$,
where $\lambda$ and $\lambda'$ are the helicities of the gluons
and $\sigma$ that of the (massless) quark line, can be written as

\begin{equation}\label{tree1}
{\cal A}_{++,+} \ = \ - \, 2 \, g^2 \sqrt{\frac{-u}{s}} 
   \left\{ \frac{s^2}{ut} \tau^a \tau^b \, + \, \frac{s}{t}
 \tau^b \tau^a  \right\}.
\end{equation}

\begin{equation}\label{tree2}
{\cal A}_{--,+} \ = \ 2 g^2 \sqrt{\frac{-u}{s}} 
   \left\{ \frac{s}{t} \tau^a \tau^b \, + \, \frac{u}{t}
 \tau^b \tau^a  \right\},
\end{equation}

\noindent
where the symbols $\tau^a$ and $\tau^b$ denote the colour matrices.
By parity conservation, we have ${\cal A}_{--,-}={\cal A}_{++,+}$
and ${\cal A}_{++,-}={\cal A}_{--,+}$.

At one-loop order, the exchanges of a vector boson of mass $m$ and
coupling $g_+$ to right-handed quarks make corrections to these matrix
elements that can be written in the forms: 

\begin{eqnarray}
\Delta {\cal A}_{++,+} & = &
 \frac{g^2 g_+^2}{8 \pi^2}\sqrt{\frac{-u}{s}} \Bigg[
  \left\{ 2 \,\left[ 1 \, -\frac{s}{t} \, -\frac{t}{s} \right]
  \left(1+\frac{m^2}{t}\right)^2 \ct0
  \right.  \nonumber \\ & &   \hspace*{-.5cm} \left.  
 - \, 2 \, \frac{t^2}{u \, s} \left(1+\frac{m^2}{t}\right)^2 \fnu
 \right.   \nonumber \\ & &   \hspace*{-.5cm} \left. 
 + \left[ 4 \frac{s}{u} \, + \, 2 \left(1-\frac{m^2}{u}\right)^2 \, 
   - \, \frac{s}{u} \left(1+\frac{m^2}{u}\right)^2 \right]
  \ln\left(1-\frac{u}{m^2} \right)
 \right.  \nonumber \\ & &   \hspace*{-.5cm} \left.  
+ \, \left[ \frac{s}{t}-2\left(1-\frac{s}{t}\right)
   \left(1+\frac{m^2}{t}\right) \right] \ln\left(\frac{-t}{m^2}\right)
- \frac{7}{2} \frac{s}{t}-\frac{5}{2}\frac{s}{u} -\frac{m^2 \, s}{t^2}
    \left(1+\frac{s^2}{u^2} \right)
\right\}  \tau^a \tau^b 
   \nonumber \\ & &   \hspace*{-.3cm} + \ \left\{ 
- \, 2 \frac{s^2}{u \, t}\left(1+\frac{m^2}{t}\right)^2 
  \ct0
  \right.  \nonumber \\ & &   \hspace*{-.5cm} \left. 
- \, 2 \, \frac{s}{u} \left(1-\frac{m^2}{s}\right)^2 \fns
  \right.  \nonumber \\ & &   \hspace*{-.5cm} \left.  
 + \left[ 2 \frac{m^2}{t}-2 \frac{m^2 \, s}{t^2} - 3 \frac{s}{t} \right]
 \ln\left(\frac{-t}{m^2}\right)
 + \, \frac{7}{2} \frac{s}{t} 
  \, - \, 2 \frac{m^2}{t} \left(1-\frac{s}{t}\right)
\right\} \tau^b\tau^a
\Bigg],  \label{amp1} \end{eqnarray}

\begin{eqnarray}
\Delta {\cal A}_{--,+} & = &
 \frac{g^2 g_+^2}{8 \pi^2}\sqrt{\frac{-u}{s}} \Bigg[ \left\{
  2\frac{u}{t} \left(1+\frac{m^2}{t}\right)^2
  \ct0
  \right.  \nonumber \\ & &   \hspace*{-.5cm}  \left. 
+ \,  2 \, \left(1-\frac{m^2}{u}\right)^2 \fnu
  \right.  \nonumber \\ & &   \hspace*{-.9cm} \ \left. 
 + \left[ 3\frac{s}{t} \, + \, 2 \frac{m^2}{u} \, - \,
  2 \frac{u \, m^2}{t^2} \right]
 \ln\left(\frac{-t}{m^2}\right)- \frac{7}{2} \frac{s}{t} 
  \, - \, 2 \frac{m^2}{u} \left(1-\frac{u^2}{t^2}\right)
\right\} \tau^a \tau^b 
   \nonumber \\ & &   \hspace*{-.3cm} + \   \left\{
 - \, 2 \, \left( \frac{u}{t} \, + \,  \frac{t^2}{u^2}\right)
    \left(1+\frac{m^2}{t}\right)^2  \ct0
  \right.  \nonumber \\ & &   \hspace*{-.5cm} \left.  
 + \, 2 \, \left(\frac{(t+m^2)}{u} \right)^2 \fns
  \right.  \nonumber \\ & &   \hspace*{-.5cm} \left.  
 - \, \left[ 4 \, + \,  2 \, \frac{s}{u} \left( 1-\frac{m^2}{s} \right)^2
 - \left( 1+\frac{m^2}{s}  \right)^2 \right]
  \ln\left( 1-\frac{s}{m^2} \right)
  \right.  \nonumber \\ & &   \hspace*{-.5cm} \left.  
 + \, \left[ 2 \frac{s}{u} \left(1-\frac{m^2}{s} \right) \, - \, 3 \,
  \frac{s}{t} \, + \, 2 \frac{u \, m^2}{t^2} 
  \right] \ln\left(\frac{-t}{m^2}\right)
 \right.  \nonumber \\ & &   \hspace*{4.cm} \left.  
 + \, \frac{7}{2} \frac{s}{t} \, + \, \frac{5}{2} \,  + \, \frac{m^2}{s}
    \left(1 \, - \, 2 \frac{u \, s}{t^2} \right)
\right\} \tau^b\tau^a
\Bigg].  \label{amp2} \end{eqnarray}

\noindent
We observe that $\Delta {\cal A}_{--,-}$ can be obtained from $\Delta
{\cal A}_{++,+}$ and $\Delta {\cal A}_{++,-}$ from $\Delta {\cal
A}_{--,+}$ by replacing $g_+$ with the coupling of the gauge boson to
left-handed quarks, $g_-$. 

In the Abelian limit, $\tau^a \tau^b =\tau^b \tau^a$, and after
rearranging the couplings appropriately, we recover the results
of~\cite{DD} for the $Z-$boson corrections to polarized Compton
scattering.

\section{Parton-level asymmetries}
In this section, we display the parity-violating
asymmetries of the cross sections for the
basic partonic subprocesses
\beq q \, + \, q \ \to \ q \, + \, q , \label{proc1}  \eeq
\beq  q \, + \bar{q} \to   q \, + \bar{q} , \label{proc2} \eeq
\beq  q \ + \, g \ \to \ q \, + \, g, \label{proc3} \eeq 
\beq q \, + \bar{q} \to   g \, + g, \label{proc4} \eeq
\beq  g \, + g \to q \, + \bar{q}. \label{proc5} \eeq
For the sake of illustration, we choose $q=d$.
The processes (\ref{proc1}) and (\ref{proc2})  
can occur at tree level via $W-$ or $Z-$boson
exchange, and all their EW and QCD ingredients have been calculated  
in~\cite{bgs}. The processes (\ref{proc3})--(\ref{proc5}) occur
only at the one-loop level, and are computed numerically here using the
amplitudes given in the previous section.

The parity-violating asymmetries of interest are~\cite{Bunce}:
\beq A_L \, d\sigma \ \equiv \ d\sigma_- \, - \, d\sigma_+, \eeq
for the case where only one incident beam is polarized, and
 \beq A_{PV} \, d\sigma \ \equiv \ d\sigma_{--} \, - \, d\sigma_{++}, \eeq
when both incident beams are polarized\footnote{In the case of
 quark-gluon scattering, the first index refers to the helicity of the 
quark.}. 

\FIGURE{
\centerline{\epsfig{file=PV0.ps, width=11cm, angle=90}}
\caption{Parton-level asymmetries at rapidity  $y=0$, as functions
of the transverse momentum $p_T$.}
\label{fig2}}

In Fig.~\ref{fig2} we show these asymmetries in the central rapidity
region.  The dot-dashed line is the quark-quark scattering contribution
calculated in~\cite{bgs}, and the dashed line refers to quark-gluon
scattering. We see that this asymmetry has the opposite sign from the
quark-quark contribution, except for $A_L$ in a small region near the weak
threshold $p_T \sim M_W/2$. The magnitude of the effect increases with
transverse momentum, reaching about 10\% at the upper end of the spectrum.
The dotted line refers to the crossed process of quark-antiquark
annihilation into two gluons and we see that in this case the asymmetry is
smaller by two orders of magnitude. 

\FIGURE{ 
\centerline{\epsfig{file=PV1.ps, width=11cm, angle=90}}
\caption{Parton-level asymmetries at rapidity  $y=1$, as functions
of the transverse momentum $p_T$.}
\label{fig3}}

In Fig.~\ref{fig3} we show the same plots, but now at relatively large
rapidity, $y=1$.  Here we see that the contributions from quark-gluon
scattering to both asymmetries can have either sign, and at sufficiently
large transverse momenta the contributions to both the asymmetries are
comparable in magnitude, and opposite in sign, to that from tree-level
quark-quark scattering. Also shown in Fig.~\ref{fig3} is the other crossed
process, namely gluon-gluon annihilation into a quark-antiquark pair. 
This was suppressed in Fig. \ref{fig2}, because it is negligibly small for
small rapidity. Once again, we see that the asymmetry is two orders of
magnitude smaller than in the quark-gluon channel. 

The very strong suppression of both crossed channels
(\ref{proc4})--(\ref{proc5}), compared to the subprocess (\ref{proc3}),
irrespective of the rapidity, was not intuitively obvious to us.  It is in
part responsible for the smallness of the overall corrections at the
hadron level, discussed below. Finally, we note the overall normalization
of all the curves, namely nanobarn/GeV, which takes into account the
relative weights of the various subprocesses in the integrated quantities.

\section{Asymmetries in polarized proton-proton scattering}

In this section we show the same asymmetries, but now in polarized
proton-proton scattering, i.e., after folding the parton-level cross
sections with polarized parton distribution functions. As examples, we
take the latter from the LO set A of Ref.~\cite{gs} (GS) 
(see also \cite{grv94})\footnote{The parton distribution functions are 
here evaluated at the scale
$Q=\sqrt{\hat{s}}$, i.e., the centre-of-mass (CM) energy at the parton
level. The same choice is made for the argument of $\alpha_s$.} at two
collider energies, $\sqrt s= 300$ and 600 GeV, representative of the
RHIC-Spin programme. As already mentioned, one might {\sl a priori} have
thought that the effect at the parton level discussed in the preceding
section could be enhanced by a substantial contribution due to the
asymmetry in the polarized gluon distribution function. In fact, we show
that this is not the case.  Qualitatively, one can understand this effect
from the fact that the asymmetries at the parton level were only
significant at large values of transverse momentum and/or rapidity. These
kinematic regions probe the large $x$ spectrum of the distributions, where
the gluon content of the proton is no longer significant.

\FIGURE{
\centerline{\epsfig{file=hadron0.ps, width=11cm, angle=90}}
\caption{Differential cross sections and asymmetries in polarized
proton-proton scattering at rapidity  $y=0$, as functions
of the transverse momentum $p_T$.}
\label{fig4}}

In Fig. \ref{fig4} we present the differential cross sections and
asymmetries for zero rapidity.  Beneath each plot, we also show the
percentage contribution due to the one-loop weak corrections, normalised
to the LO rate.  Each graph has two sets of curves, corresponding to the
two mentioned CM energies, as indicated.

The top-left graph is the total differential cross-section, which is
dominated by LO QCD. Here, as expected, the one-loop weak corrections are
of the order of 0.1\%. The top-right graph shows another asymmetry,
$A_{LL}$, defined as

\beq A_{LL} \, d\sigma \ \equiv \ d\sigma_{++} \, - \, d\sigma_{+-} \, + \,
d\sigma_{--} \, - \, d\sigma_{-+} \,
, \eeq
which is also present in QCD and depends entirely on the asymmetry of the
parton distributions.  Here we note that the contributions from the
one-loop weak corrections are even smaller than for the differential
cross-section itself, at both energies.

The lower two graphs show the parity-violating asymmetries, normalized to
the total cross section, so that the dimensionless quantities $A_L$ and
$A_{PV}$ are shown.  We have normalised $A_{LL}$ in the same way.  We see
that the effect of the weak corrections to quark-gluon scattering is
generally to reduce the two asymmetries by a few percent, i.e., of the
same order as was found at the parton level. 

\FIGURE{ 
\centerline{\epsfig{file=hadron1.ps, width=11cm, angle=90}}
\caption{Differential cross sections and asymmetries in polarized 
proton-proton scattering at rapidity  $y=1$, as functions
of the transverse momentum $p_T$.}
\label{fig5} }

Fig.~\ref{fig5} plots the same quantities as in Fig.~\ref{fig4}, but now
at rapidity $y=1$. Here is where one gets the largest effects at the
one-loop level, as the weak corrections can reduce the parity-violating
asymmetries by up to about 10--12\%. The effects on the total rate and on
$A_{LL}$ are of the same order as at lower rapidity.

\FIGURE{ 
\centerline{\epsfig{file=integral.ps, width=11cm, angle=90}}
\caption{Differential cross-sections and asymmetries in polarized
proton-proton
scattering integrated over the rapidity range $|y|<1$, as functions
of the transverse momentum $p_T$.}
\label{fig6}}

Finally, in Fig.~\ref{fig6}, we display the usual selection of curves, but
now integrated over rapidity, from $y=0$ to $|y|=1$. The shapes modulate
between the extreme trends seen in the two previous figures. The salient
features already appreciated at fixed rapidities, that one-loop weak
corrections are larger at higher energy and carry some visible structure
induced by the $W-$ or $Z-$mass threshold effects, remain clearly visible. 

\section{Summary and conclusions}

We have calculated the one-loop weak corrections to the quark-gluon
scattering differential cross section, as a function of the transverse
momentum and for selected values of rapidity, and crossed channels. We
have evaluated the corresponding contributions to parity-violating
asymmetries in polarized proton-proton scattering at energies relevant to
RHIC-Spin.  We have found that in certain kinematic regions these
higher-order effects can be as large as 10\% of the tree-level
contributions to some parity-violating asymmetries.  We recall that the
study of these asymmetries has gathered particular attention
lately~\cite{Bunce}, both because they are sensitive to the
helicity-dependent components of the parton distribution functions, and
because they can carry the distinctive hallmark of new physics. 

It turns out that, the larger the rapidity and the collider energy, the
bigger the effect of the one-loop weak corrections, with some resonant
enhancement in the transverse momentum spectrum in the vicinity of the
$W-$ and $Z-$mass thresholds. These corrections are dominated by the
contributions induced by quark-gluon scattering, as both quark-antiquark
annihilation into two gluons and gluon-gluon scattering to a
quark-antiquark pair are considerably suppressed in comparison. In fact,
at the parton level, one-loop corrections via quark-gluon scattering can
be comparable to the tree-level contribution due to electroweak processes
involving four external (anti)quarks. 

These parton-level effects are not enhanced when one folds the
parton-level cross sections with polarized distribution functions. The
main reason is that the effects on the parity-violating asymmetries are
only large for large values of the Bjorken $x$ variable, where the gluon
distribution function is generally small. 

In order to check whether our results depend on the particular polarized
distribution functions we used, we have repeated our exercises using
different sets~\cite{GRSV}, and find no significant variations. We find
that the central qualitative result, namely that the contribution from
quark-gluon scattering contributes around 10\% to the parity-violating
asymmetries $A_L$ and $A_{PV}$, is unchanged. Nevertheless, we note that
at the higher CM energy of 600 GeV, there is some sensitivity to the
choice of polarized parton distributions.  For the central rapidity case,
the magnitude of the quark-gluon contribution is diminished by around 2\%
if one uses the polarized distribution functions of GRSV~\cite{GRSV} or
set C of GS~\cite{gs}.  For larger
rapidity, $y=1$, the GSRV distributions and sets A,B of the GS
distributions give very similar results, whereas set C of GS leads to
substantial enhancements of the parity-violating asymmetries, namely up to
$-25\%$, in the region of the mass threshold $p_T \sim M_W$.
Set C of~\cite{gs} differs qualitatively from the other
sets in that the contribution to the proton spin from the gluons is {\sl
negative} over a substantial range of $x$.

Our calculations can readily be extended to evaluate similar corrections
to direct photon production, $\qrd +~g \to \qrd +~\gamma$ and $\bar{q} + q
\to g + \gamma$. Besides these processes of potential interest to
RHIC-Spin, the generalization to the case of three-jet production in $e^+
e^-$ collisions, namely the one-loop weak corrections to $\gamma^* / Z^*
\to q~+~\bar{q}~+~g$, is also relatively straightforward.  This is
especially worthy of attention, given that the NNLO QCD corrections to
this process are also expected to become available soon, and in view of
the copious three-jet data samples furnished by LEP.  Presumably though,
the one-loop weak corrections will be suppressed at the $Z^0$ peak by a
factor $G_F Q^2$, where $Q$ is a typical virtuality scale of order
$M_{Z^0}$.  Hence, they would be relatively more important at LEP~2 or at
a higher-energy $e^+ e^-$ linear collider. 

\section*{Acknowledgements}

We are grateful to Dima Bardin,  Stefano Catani, Stefan Dittmaier 
and Thomas Gehrmann for several fruitful discussions.

\end{document}